\documentclass[11pt, a4paper]{article}
\usepackage{setspace} 
\usepackage{amsmath}
\usepackage{color}
\usepackage{graphics}
\usepackage[round,numbers,sort&compress]{natbib}
\newcommand{\df}[2]{\frac{\partial #1}{\partial #2}} 
\newcommand{\bsigma}{{\boldsymbol\sigma}}

\newcommand{\uA}{\mu\mathrm{A}}  
\newcommand{\cm}{\mathrm{cm}}    
\newcommand{\m}{\mathrm{m}}      
\newcommand{\ms}{\mathrm{ms}}    
\newcommand{\us}{\mu\mathrm{s}}  
\newcommand{\uF}{\mu\mathrm{F}}  
\newcommand{\mV}{\mathrm{mV}}    
\renewcommand{\S}{\mathrm{S}}    

\newcommand{\eg}[1]{{\it e.g.\/}\ifx#1.\else\expandafter#1\fi}     
\newcommand{\etal}[1]{{\it et al.\/}\ifx#1.\else\expandafter#1\fi}  
\newcommand{\ie}[1]{{\it i.e.\/}\ifx#1.\else\expandafter#1\fi}      

\newcommand{\Fig}[1]{Fig.~\ref{fig:#1}}
\newcommand{\fig}[1]{Fig.~\ref{fig:#1}}

\newcommand{\myfigure}[3]{
\begin{figure}[t]
\centerline{\includegraphics{#1.eps}}
\caption{#2}
\label{fig:#3}
\end{figure}
}

\title{Low energy defibrillation in human cardiac tissue: a simulation study}

\author{
  Stuart~W.~Morgan, \\ Department of Mathematical Sciences, University of Liverpool, UK \\
\and
  Gernot~Plank, \\ Institute of Biophysics, Medical University of Graz, Austria \\
\and 
  Irina~V.~Biktasheva, \\ Department of Computer Science, University of Liverpool, UK \\
\and
  Vadim~N.~Biktashev\thanks{
           Corresponding author.  Address: 
           Department of Mathematical Sciences,
	   University of Liverpool,
	   Liverpool, L69 7ZL, UK,
	   Tel.:~+44-151-7944006, Fax:~+44-151-7944061} \\
  Department of Mathematical Sciences,
  University of Liverpool, UK
}

\date{Submitted: September 22, 2008 \\ Final version: November 19, 2008}

\begin{document}
\maketitle
\thispagestyle{empty}

\begin{abstract}
  We aim to assess the effectiveness of feedback controlled resonant
  drift pacing as a method for low energy defibrillation.

  Antitachycardia pacing is the only low energy defibrillation
  approach to have gained clinical significance, but it is still
  suboptimal. Low energy defibrillation would avoid adverse side
  effects associated with high voltage shocks and allow the
  application of ICD therapy where it is not tolerated today.

  We present results of computer simulations of a bidomain model of
  cardiac tissue with human atrial ionic kinetics. Re-entry was
  initiated and low energy shocks were applied with the same period as
  the re-entry, using feedback to maintain resonance.  We demonstrate
  that such stimulation can move the core of re-entrant patterns, in
  the direction depending on location of electrodes and a time delay
  in the feedback. Termination of re-entry is achieved with shock
  strength one order of magnitude weaker than in conventional
  single-shock defibrillation.

  We conclude that resonant drift pacing can terminate re-entry at a
  fraction of the shock strength currently used for defibrillation and
  can potentially work where antitachycardia pacing fails, due to the
  feedback mechanisms. Success depends on a number of details which
  these numerical simulations have uncovered.

\end{abstract}


\clearpage
\pagestyle{myheadings}
\pagenumbering{arabic}
\setcounter{page}{1}
  
\section*{Introduction}

Several clinical trials established that the timely application of an
electric shock, particularly with ICD, is the only reliable therapy to
prevent sudden cardiac death \cite{ClinTrial2}.  However, the strong
shocks required are reported to have serious adverse effects, most
prominently via electroporation, alterations of the action potential
waveform and duration \cite{Waveform1}, depolarization of the resting
potential \cite{DeBruin99_rest}, increased pacing thresholds
\cite{sambelashvili04:_threshold}, loss of excitability
\cite{Sharma05}, and transient ectopic afterdepolarizations which may
initiate postshock arrhythmias \cite{zivin99:_postdefib}.  Other
studies reported mechanical aftereffects such as mechanical
dysfunction (stunning), increases in contractility \cite{MechDis1} and
hemodynamically mediated symptoms \cite{tokano98:_hemodynamics}.
Psychological effects on patients play a non-negligible role
\cite{boriani05:_mechanisms_pain}, and despite long-term survival
benefits, patients suffering from arrhythmias which are not
immediately life-threatening do not tolerate ICD therapy
\cite{Tolerance}.  Biphasic \cite{walcott95:_choose}, multiphasic
\cite{zhang06:_quadriphasic}, and truncated exponential
\cite{TruncExp} shock waveforms defibrillate at a lower threshold, but
still too high for painless defibrillation.

Several approaches to minimize defibrillation energy by employing
smarter protocols are under examination, but so far only
antitachycardia pacing has gained clinical significance.
Antitachycardia pacing is a series of weak shocks applied at a
frequency higher than the intrinsic frequency of the arrhythmia.  This
therapy has conventionally been applied to slower, presumably
hemodynamically tolerated, ventricular tachycardias. Fast ventricular
tachycardias ($188$ to $250$~beats/min) typically receive high
amplitude shock therapy, even though antitachycardia pacing may work
\cite{ATP2}.  Although the mechanisms responsible for antitachycardia
pacing failure are not fully understood, an inherent weakness is
evident: a fixed pacing frequency is likely to be suboptimal depending
on the arrhythmia.  No arrhythmia-specific input is used to form or
adjust the antitachycardia pacing sequence.

In this study we assume that arrhythmias are sustained by re-entry and
consider a method which employs feedback-driven pacing to control and
eliminate the re-entry cores, by moving them until they hit
inexcitable obstacles or each other, and annihilate. The method relies
on a phenomenon of ``resonant drift'' \cite{%
  Davydov-etal-1988,%
  Agladze-etal-1987%
}: the drift of re-entrant waves when periodic, low energy, shocks are
applied in resonance with the period of the re-entry. A feedback
algorithm \cite{Biktashev-Holden-1994} is used to maintain the
resonance.  Resonant drift and its feedback control, herein referred
to as resonant drift pacing, have only been studied experimentally in
the Belousov-Zhabotinsky reaction medium and in simulations of
simplified models of cardiac tissue (\eg~\cite{%
  Zykov-Engel-2007,%
  Biktashev-Holden-1996,%
  Panfilov-etal-2000%
}).

The goal of the present simulation study is to investigate the
effectiveness of using resonant drift pacing for low-voltage
defibrillation. Previous studies of resonant drift were in models
which were very different from modern models of cardiac tissue in many
important respects, and their relevance for low-voltage defibrillation
is debatable.  Here we use an anisotropic bidomain model of cardiac
tissue with microscopic heterogeneities and realistic cellular
kinetics of human atria.  Shocks are applied by injecting current
into, and withdrawing from, the extracellular space. Such description
was used in simulations before, but only to study high-voltage
single-shock defibrillation \cite{Plank-etal-2005}

Our results show that in this model setting, resonant drift pacing can
be used to move the core of re-entrant activation
patterns. Termination can be achieved with high probability, and
within the time deemed acceptable by clinicians for antitachycardia
pacing to work, at a fraction of the conventional single shock
defibrillation strength, by moving the cores until they hit an
anatomical boundary, or annihilate with each other. We show that in a
realistically anisotropic model, direction of movement depends on
electrode location and time delays of the shock application, which is
in agreement with previous studies \eg~\cite{%
  Davydov-etal-1988,%
  Biktashev-Holden-1996%
}. Knowing the electrode location and the anatomy of the heart, the
best delay could be estimated to move the core in the direction of a
suitable anatomical structure or boundary (i.e. an inexcitable piece
of tissue) which is likely to terminate the re-entrant wave.

\section*{Methods}

\subsection*{Governing equations}

The bidomain model of cardiac tissue is most widely used to study
defibrillation-related phenomena~\cite{Tray}.  The system can be
written as
\begin{eqnarray}
C_{m}\df{V}{t} 
  & = & -I_{ion} 
  + \frac{1}{\beta} \nabla \cdot \left({\bsigma_{i}}\nabla V \right)
  + \frac{1}{\beta} \nabla \cdot \left({\bsigma_{i}}\nabla \phi_{e} \right), \label{para}\\
\nabla \cdot \left( \left({\bsigma_{i}+ \bsigma_{e}}\right)\nabla
  \phi_{e} \right) 
  & = & -\nabla \cdot \left( {\bsigma_{i}} \nabla V \right) 
  + I_{e}, \label{ell}
\end{eqnarray}
where $\phi_i$ and $\phi_e$ are the intracellular and extracellular
potential distributions, $\beta$ is the average cell's surface to
volume ratio, $C_{m}$ is the membrane capacitance per unit area, $V =
\phi_{i} - \phi_{e}$ is the transmembrane potential, ${\bsigma_{i}}$
and ${\bsigma_{e}}$ are the intracellular and extracellular
conductivity tensors respectively, $I_{e}$ is the extracellular
current, and $I_{ion}$ is the ionic current density through the
membrane.

In absence of extracellular current $I_{e}$, if variations of
$\phi_{e}$ are negligible compared to $\phi_{i}$, or if the anisotropy
ratios in both domains are the same, then (\ref{para}) and (\ref{ell})
can be replaced by a simpler monodomain equation,
\begin{eqnarray}
C_{m}\df{V}{t} 
  & = & -I_{ion} 
  + \frac{1}{\beta} \nabla \cdot \left({\bsigma_{m}}\nabla V \right),
\end{eqnarray}
where $\bsigma_{m}$ is the monodomain conductivity tensor.

We used a bidomain description of cardiac tissue, with the exception
of monodomain description later than $10\,\ms$ after shocks when
determining single-shock defibrillation thresholds.  A recent
comparison of the monodomain and bidomain models suggests that they
yield very similar results as long as no strong electric fields are
applied \cite{potse06:_comparison}.  To allow comparison with
\cite{Plank-etal-2005}, we have used the Courtemanche \etal\ human
atrial model \cite{courtemanche98:_human} for the ionic currents, with
the alterations described below.  This model is well established and
very detailed, taking into account all major ionic transport
mechanisms and intracellular calcium handling.  Note that resonant
drift in monodomain models was investigated both with atrial
\cite{Biktashev-Holden-1995a} and ventricular
\cite{Biktashev-Holden-1996} cellular kinetics and results were
similar, modulo the difference between steadily rotating and
meandering spirals. So we expect that our present result should be
interesting for ventricular fibrillation too, subject to a proper
account of other important differences between ventricles and atria.

\subsection*{Numerical methods and parameters}

All simulations were performed by the Cardiac Arrythmia Research
Package (CARP) \cite{CARP,VigCARP,plank07:_amg}. We used a numerical
setup similar to Plank \etal\ \cite{Plank-etal-2005}.  We used a thin
sheet of cardiac tissue $4 \times 2 \times 0.02\,\cm^{3}$ with the
fibres along the $x$ axis, no-flux boundary conditions and no
surrounding bath.

As in \cite{Plank-etal-2005}, the intracellular conductivities in
bidomain calculations were fluctuating, with conductivities at
different points being uncorrelated random numbers within $\pm50\%$ of
the average,
\begin{eqnarray}
\sigma_{ij} & = & \bar{\sigma}_{ij}(1+F\eta),
\end{eqnarray}
where $j=x,y,z$, the level of fluctations was fixed to $F=0.5$, and
$\eta \in [-1,1]$ are independent equidistributed random numbers.  The
justification for introducing such fluctuations is that although
cardiac tissue at a macroscopic scale is frequently approximated as a
homogeneous bisyncytium, this is not valid at a microscopic scale, and
this makes an essential difference when an external electric field is
applied. As the simulation study \cite{Plank-etal-2005} showed, the
presence of fluctuations was fundamental for the mechanism of
defibrillation, although the exact value of $F$ was less
important. Apart from the $F=0.5$ which was the maximal considered in
\cite{Plank-etal-2005}, we also tried $F=0.25$ in a few test
simulations and found the difference not principal.

\begin{table}[tp]
  \centering
  \begin{tabular}[c]{|l|l|}\hline
    Object/quantity & Notation/value \\\hline
    Bidomain model &
    \begin{minipage}{0.5\linewidth}\small
\begin{eqnarray*}
&& \hspace*{-1em} C_{m}\df{V}{t} = -I_{ion} + \frac{1}{\beta} \nabla \cdot
   \left( \bsigma_{i} \nabla ( V  + \phi_{e}) \right), \\
&& \hspace*{-1em} \nabla \cdot \left( \left(\bsigma_{i}+ \bsigma_{e}\right)\nabla \phi_{e} \right) = -\nabla \cdot \left( \bsigma_{i} \nabla V \right) + I_{e}
\end{eqnarray*}
    \end{minipage} \\\hline
    Monodomain model &
    \begin{minipage}{0.5\linewidth}\small
\begin{eqnarray*}
&& C_{m}\df{V}{t} = -I_{ion} + \frac{1}{\beta} \nabla \cdot \left(\bsigma_{m}\nabla V \right)
\end{eqnarray*}
    \end{minipage} \\\hline
Intra- and extra-cellular potentials & $\phi_i$  and $\phi_e$ \\\hline
Transmembrane potential & $V = \phi_{i} - \phi_{e}$ \\\hline
Transmembrane ionic current density & $I_{ion}$, as in \cite{courtemanche98:_human}\mbox{}, with alterations \\\hline
External extracellular current density & $I_{e}(x,y,z,t)$, by stimulating electrodes \\\hline
Average cell's surface to volume ratio & $\beta=1400\,\cm^{-1}$ \\\hline
Membrane capacitance per unit area & $C_{m}=1.0\,\uF/\cm^{2}$ \\\hline
Extracellular conductivity, tensor & $\bsigma_{e}$ \\\hline
\dots, along the fibers & $\sigma_{ex}=0.625 \,\S/\m$ \\\hline
\dots, across the fibers & $\sigma_{ey}=\sigma_{ez}=0.236 \,\S/\m$ \\\hline
Intracellular conductivity, tensor &
                        $\bsigma_{i} = \bar{\bsigma}_{i}(1+F\eta(x,y,z))$\\\hline
\dots, fluctuations intensity & $F=0.5$ \\\hline
\dots,  
uncorrelated equidistributed noise & $\eta(x,y,z) \in [-1,1]$\\\hline
\dots, average, along the fibers & $\bar\sigma_{ix}=0.174\,\S/\m$ \\\hline
\dots, average, across the fibers & $\bar\sigma_{iy}=\bar\sigma_{iz}=0.019 \,\S/\m$ \\\hline
Monodomain tissue conductivity, tensor & $\bsigma_{m}$ \\\hline
\dots, along the fibres & $\sigma_{mx}=0.146 \,\S/\m$ \\\hline
\dots, across the fibres & $\sigma_{my}=\sigma_{mz}=0.0182 \,\S/\m$ \\\hline
Space discretization step & $0.01\,\cm$ \\\hline
Time discretization steps, most of the time & $10\,\us$ \\\hline
\dots, during and $10\,\ms$ after single shocks & $1\,\us$ \\\hline
  \end{tabular}
  \caption{Details of calculations}
  \label{CalcDetails}
\end{table}

Values of the numerous parameters used in this study are presented in
Table 1.

\subsection*{Visualization} \label{visualization}

\myfigure{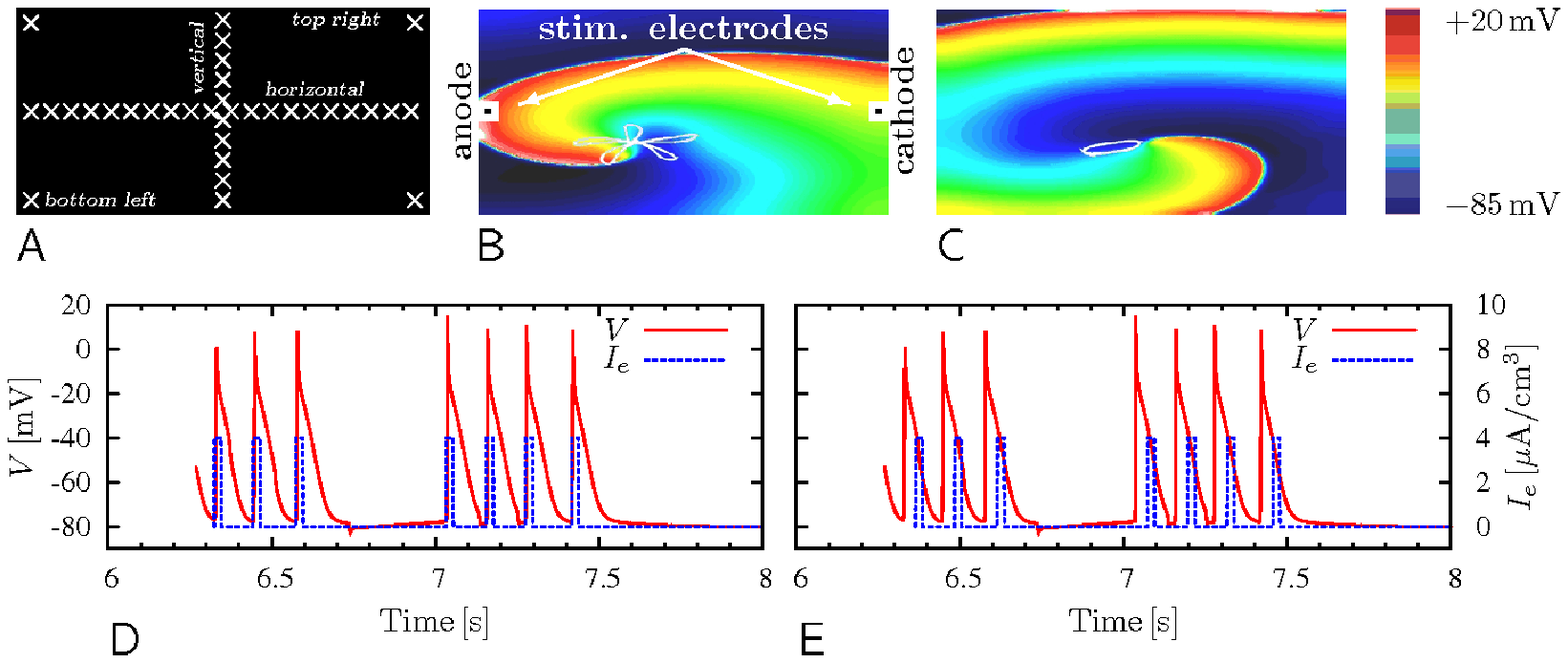}{
  A: Locations of registration electrodes.
  B:  Meandering re-entry with the trajectory of the tip (white curve) and the
  location of the stimulating electrodes.
  C: Stationary-rotating re-entry with the  trajectory of the tip. 
  Legend on the right: colour-coding of $V(x,y,t)$.
  D,E: Feedback algorithm with $t_{delay}=0\,\ms$ (D)
  and $t_{delay}=30\,\ms$ (E). 
  Red solid line: registered signal, 
  dashed blue line: stimulation shocks, $A=4 \times 10^6\,\uA/\cm^3$.
  Intervals between shocks are the same as the re-entry periods as
  registered by the electrode. 
}{figfb}

We used CARP's transmembrane voltage $V(x,y,t)$ output to visualize
the results of simulations, using the colour-coding shown in
\Fig{figfb}C.  The same information was used to detect the wave tips.
The wavefront at time $t$ was defined as the line
$V(x,y,t)=-23.75\,\mV$. Then the wave tips for that time were defined
as the intersections of the front at time $t$ with the front at time
$t-3\,\ms$.  The time delay of $t=3\,\ms$ was chosen purely
empirically.

To visualize drift, we used a ``stroboscopic'' method: we showed
positions of the tip synchronized with the signals detected by the
registration electrodes.

For meandering spirals this required further refinement, since the
stroboscopic selection of tips produced a congested picture, even
without any stimulation: so, a five-petal meandering pattern (as shown
in \fig{figfb}B) produced five clusters of tips.  With stimulation,
the tip picture becomes even more complicated and unreadable. So we
showed only every $N$-th stroboscopic tip position.  Since the
meandering patterns were affected by stimulation, we found that the
optimal value for $N$ was different from 5; we used $N=3$.

A possible alternative to the stroboscopic method is sliding averaging
of the trajectories. However, it requires careful choice of the
averaging window, which may be different in different situations, so
we found it less convenient.

\subsection*{Generation of re-entry patterns} \label{Gen}

We used two alterations of the Courtemanche human atrial model
\cite{courtemanche98:_human} of the ionic currents, $I_{ion}$:

\begin{enumerate}\renewcommand{\labelenumi}{(\roman{enumi})\ }
\item To prevent the transmembrane voltages from rising to
  non-physiological values during the defibrillation shocks, an
  electroporation current was included, see \cite{Plank-etal-2005} and
  references therein.  In addition, a formulation for a
  acetylcholine-dependent potassium current, $I_{K(ACh)}$, was added
  \cite{kneller,Plank-etal-2005}.  This ionic model was used to
  generate a meandering re-entry, see \fig{figfb}B.

\item In addition to electroporation and $I_{K(ACh)}$, we used a
  $65\%$ block of the slow inward L-type $\mathrm{Ca}^{2+}$ current
  coupled with a nine-fold increase in the slow delayed outward
  $\mathrm{K}^+$ current and the rapid delayed outward $\mathrm{K}^+$
  current as suggested by Xie \etal\ \cite{Xie}. This ionic model was
  used to generate a stationary rotating re-entry, see \fig{figfb}C.
\end{enumerate}

To initiate re-entry, we used an S1-S2 protocol.

\subsection*{Single shock defibrillation benchmark} \label{ssd}

Monophasic current shocks $I_{e}(x,y,z,t)$ were injected into, and
withdrawn from, the extracellular space via volumes $0.1 \times 0.1
\times 0.02 \,\cm^{3}$, ``stimulating electrodes'', centered along the
left and right edges of the slab (\fig{figfb}B).  The shocks were of
rectangular waveform, $5\,\ms$ duration and varied amplitude $A$.

Shocks were applied at twelve different timings $t_{0}$, separated by
$10\,\ms$ intervals, after the same initial conditions.  This covered
an entire single rotor cycle, $120\,\ms$.  A single shock was deemed
successful if no re-entry was detectable at the moment $500\,\ms$
after its end.  We define the single shock success threshold as the
shock amplitude which gives a $50$\% success rate across the twelve
timings.

\subsection*{Resonant drift pacing}

Repetitive low-amplitude shocks of the same waveform and via the same
\emph{stimulating} electrodes as above, were applied at the time
moments determined by signals received via \emph{registration}
electrodes.

Six different locations for the registration electrodes were used:
``point'' electrodes $0.02 \times 0.02 \times 0.02 \,\cm^{3}$ in the
top left, top right, bottom left and bottom right corners and ``line''
electrodes \cite{Zykov-Engel-2007} (of cross-section
$0.02\times0.02\,\cm^2$) through the whole medium either horizontally,
along the fibres or vertically, across the fibres, see \fig{figfb}A.

The signal from a registration electrode was defined as the average
potential at all the nodes covered by it. The signal \emph{triggered}
a shock application when it exceeded $-55\,\mV$. A shock was
\emph{applied} with a delay $t_{delay}$ after it had been triggered.
In most cases we set $t_{delay}=0$, see \fig{figfb}(D).  To
demonstrate its effect on the direction of the drift, we set
$t_{delay} = 30\,\ms$, about a quarter of the re-entry period, see
\fig{figfb}(E).

In all simulations, the registration electrodes were de-activated
for a ``blanking time'' $t_{blank}=50\,\ms$ after a shock application.

For the meandering re-entry, resonant drift pacing was considered
successful if the re-entry was terminated sooner than the
self-termination time of the re-entry, $16000\,\ms$. For the
stationary rotating re-entry, this time was extended to $30000\,\ms$.

\section*{Results} \label{results}

\subsection*{Single shock defibrillation results}

\myfigure{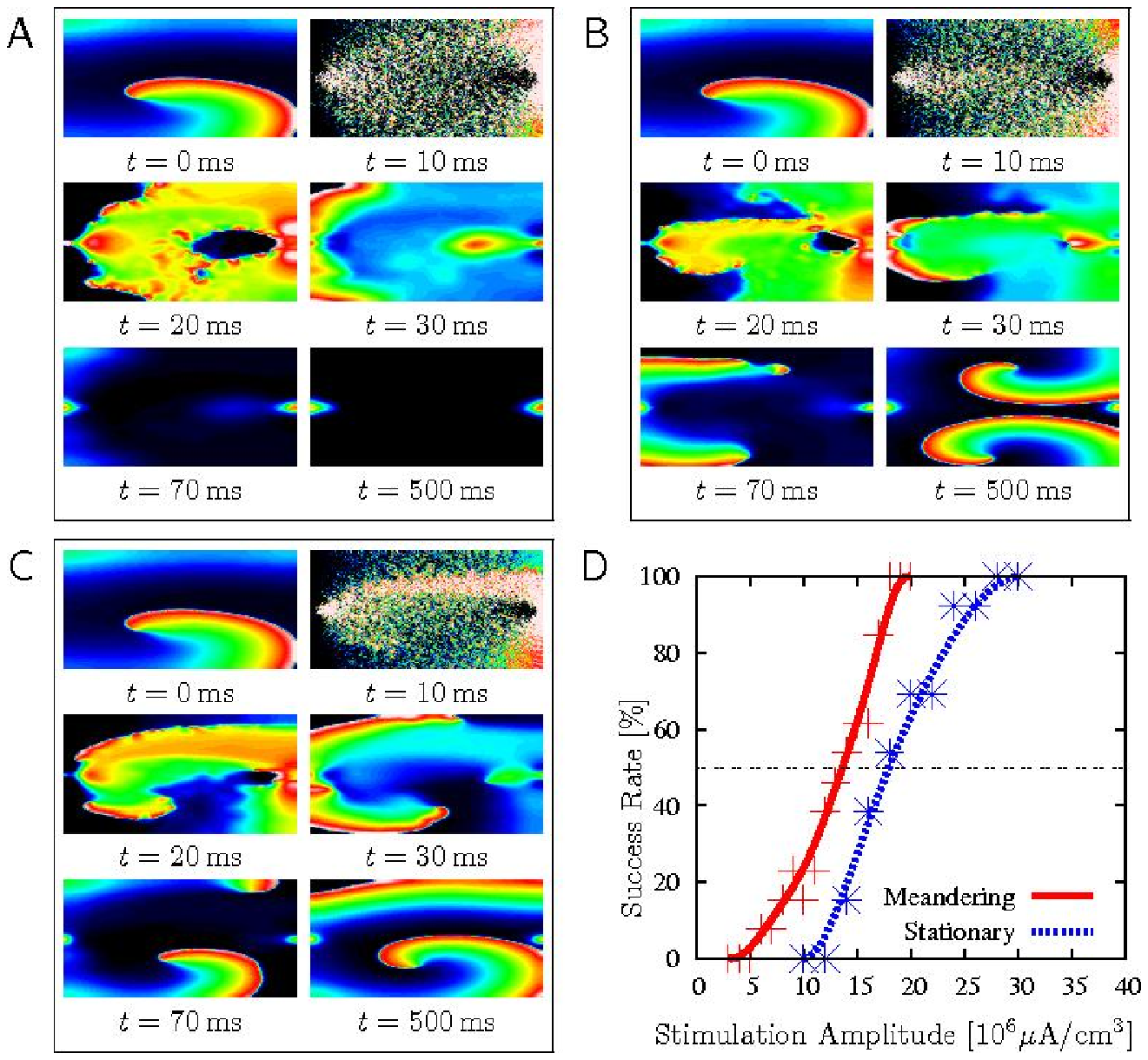}{
   Single shock defibrillation, stationary rotating re-entry.
  A: Termination, $A=16 \times 10^{6}\,\uA/\cm^3$ applied at $t=10$.
  B: Breakup, $A=14 \times 10^{6}\,\uA/\cm^3$ applied at $t=10$.
  C: Displacement, $A=12 \times 10^{6}\,\uA/\cm^3$ applied at $t=10$.
  D: Percentage of successful terminations as function of
  shock amplitude $A$. Markers: raw data;
  lines: Bezier approximation.
}{SingleShock}

We have varied the timings, and strength $A$, of the single shocks to
assess the variability of the outcomes.  Consistent with previous
observations \cite{Plank-etal-2005}, there were three typical
single-shock defibrillation outcomes for both the meandering and
stationary-rotating patterns:
\begin{itemize}
\item Strong enough shocks annihilated re-entry immediately (\fig{SingleShock}A).
\item Weaker shocks led to multiple wavebreaks (\fig{SingleShock}B).
\item Weaker still shocks only shift re-entry in space (\fig{SingleShock}C). 
\end{itemize}

We have found the single shock defibrillation threshold to be $A =
14\times 10^6\,\uA/\cm^3$ for the meandering patterns and $18\times
10^6\,\uA/\cm^3$ for the sationary-rotating patterns
(\fig{SingleShock}D).

\subsection*{Resonant drift pacing results}

We tested resonant drift pacing on the meandering, and stationary
rotating, re-entrant patterns with shock amplitudes lower than the
corresponding single shock defibrillation success thresholds.

\subsubsection*{Meandering re-entry}

\myfigure{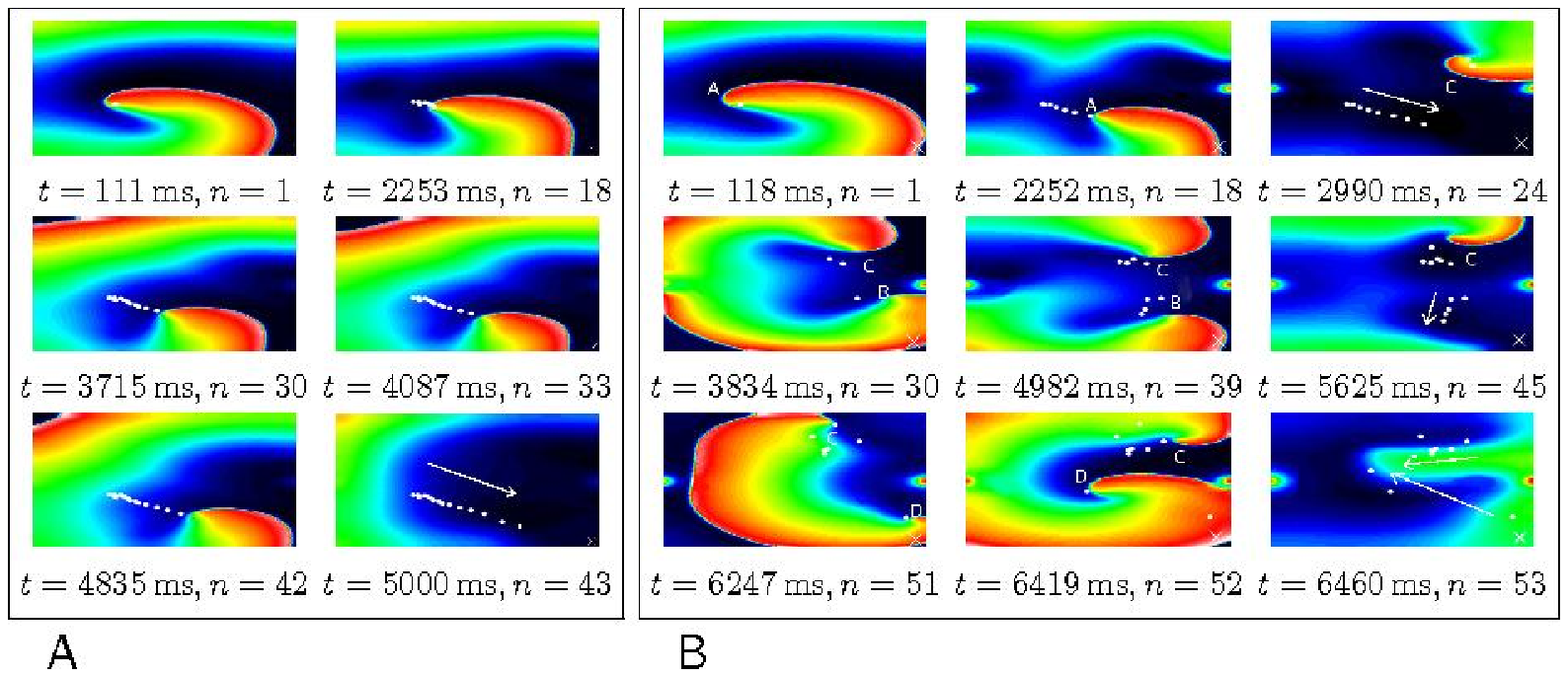}{
  Resonant drift pacing of a meandering re-entry.
  A: $A=1\times 10^6\,\uA/\cm^3$, re-entry drifts
  to the boundary and terminates. 
  Here and below, cross: location of the
  registration electrode, ``$n=\dots$'':
  number of shocks applied so far, 
  white dots: positions of the
  spiral tip at every third registered period. 
  B: $A=2\times 10^6\,\uA/\cm^3$, original re-entry ``A'' drifts to the boundary and 
  terminates. Additional re-entrant patterns ``B'', ``C'' and ``D''
  are generated by the shocks and trigger further shocks themselves. 
  ``B'' terminates at the boundary, ``C'' and
  ``D'' annihilate each other.
  \textit{NB}: resonant drift pacing handles multiple patterns.
}{MSdrift}

\paragraph{Point registration electrodes.}

Depending on the shock strength, the following outcomes were observed. 
\begin{itemize}
\item The original re-entry drifted, roughly towards the registration
  electrode, until termination on an inexcitable boundary, see
  \fig{MSdrift}A.
\item Secondary re-entrant patterns were generated near stimulating
  electrodes; both the original and newly generated re-entries
  drifted. An example is shown in \fig{MSdrift}B.  The original
  re-entry ``A'' was terminated by the first 24 shocks.  However, a
  secondary re-entrant pattern ``C'' was generated by a shock in the
  process.  After ``A'' terminated, it was ``C'' that triggered the
  subsequent shocks.  A third re-entrant pattern, ``B'' was generated
  by a further shock.  Due to the proximity of ``B'' to the
  registration electrode, it took over the control of the shock
  applications.  Re-entry ``B'' drifted to the boundary and terminated
  after a further 15 shocks.  Then ``C'' was again the solitary
  re-entry and took over the control.  A fourth re-entry, ``D'' was
  generated by a shock and, due to its proximity to the registration
  electrode, took over the control.  After 3 shocks, ``D'' collided
  with ``C'' and they annihilated each other without reaching an
  inexcitable boundary.
\end{itemize}

Note that mutual annihilation of re-entrant patterns in the bulk of
the tissue could be a plausible explanation why Pak \etal\ \cite{Pak}
saw no evidence of annihilation at the ventricular boundaries when
testing a protocol similar to the resonant drift pacing.

\paragraph{Line registration electrodes} 

\myfigure{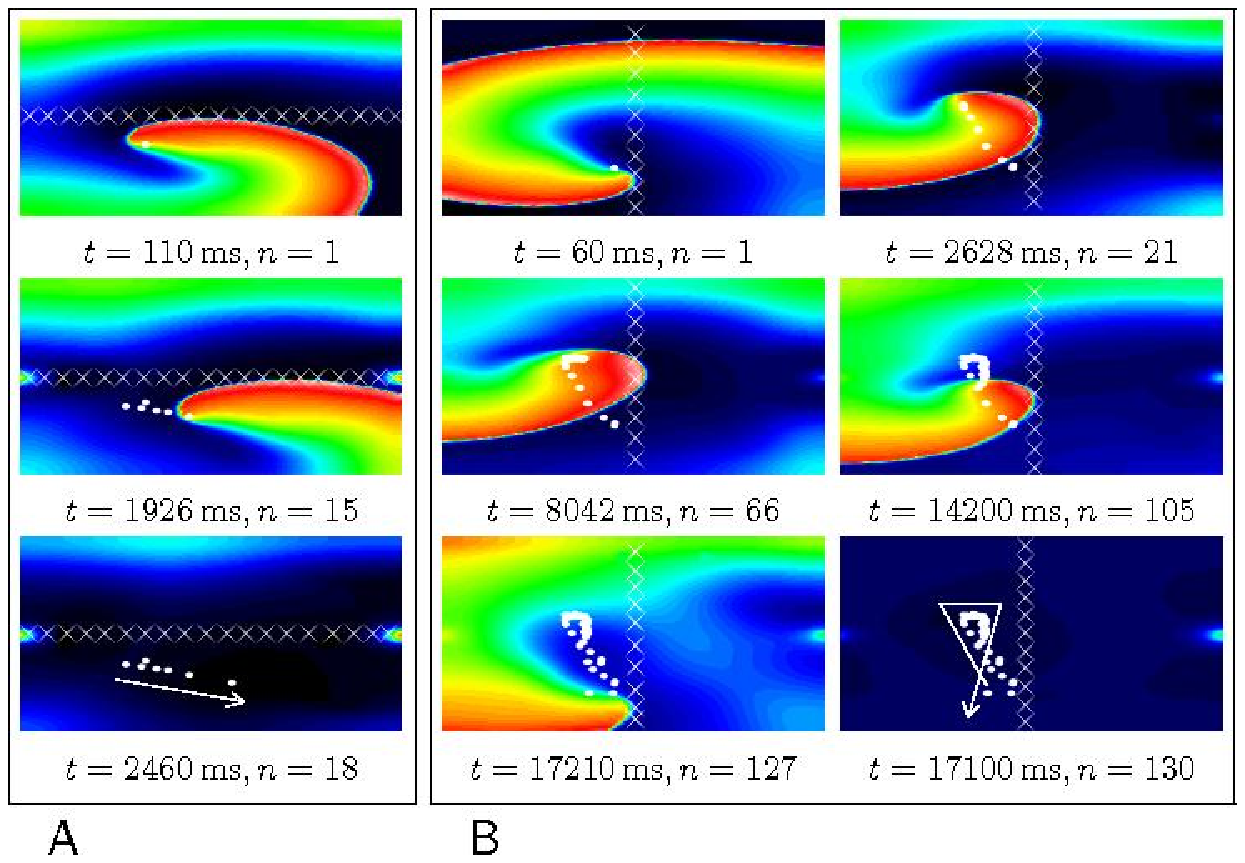}{
Resonant drift pacing with line registering electrodes.
 A: Horizontal electrode along the fibres, $A=2\times 10^6\,\uA/\cm^3$.
 The re-entry drifts resonantly at an angle to the electrode,
  reaches the boundary and terminates.
 B: Vertical electrode across the fibres, $A=1\times 10^6\,\uA/\cm^3$.
  The re-entry drifts upwards for $21$ shocks, then changes direction and
  reaches the boundary after a further $109$ shocks.
}{ZL}

\begin{itemize}
\item \emph{Horizontal electrode}: With the registration electrode
  along the fibres, the re-entry drifted at a small angle to it
  (\fig{ZL}A).
\item \emph{Vertical electrode}: With the registration electrode
  across the fibres, the drift was usually more complicated. An
  example is shown in \fig{ZL}B. The re-entry initially drifts upwards
  along the electrode. After 21 shocks, the drift turns downwards,
  until it reaches an inexcitable boundary after a further 109 shocks
  and terminates.  The trajectory of the drift crosses itself, which
  is not allowed in the asymptotic theory of drift of rigidly rotating
  spiral waves \cite{Biktashev-Holden-1994} and was not observed in
  our simulations of a stationary rotating re-entry, so we consider it
  a new feature due to meander.
\end{itemize}

\myfigure{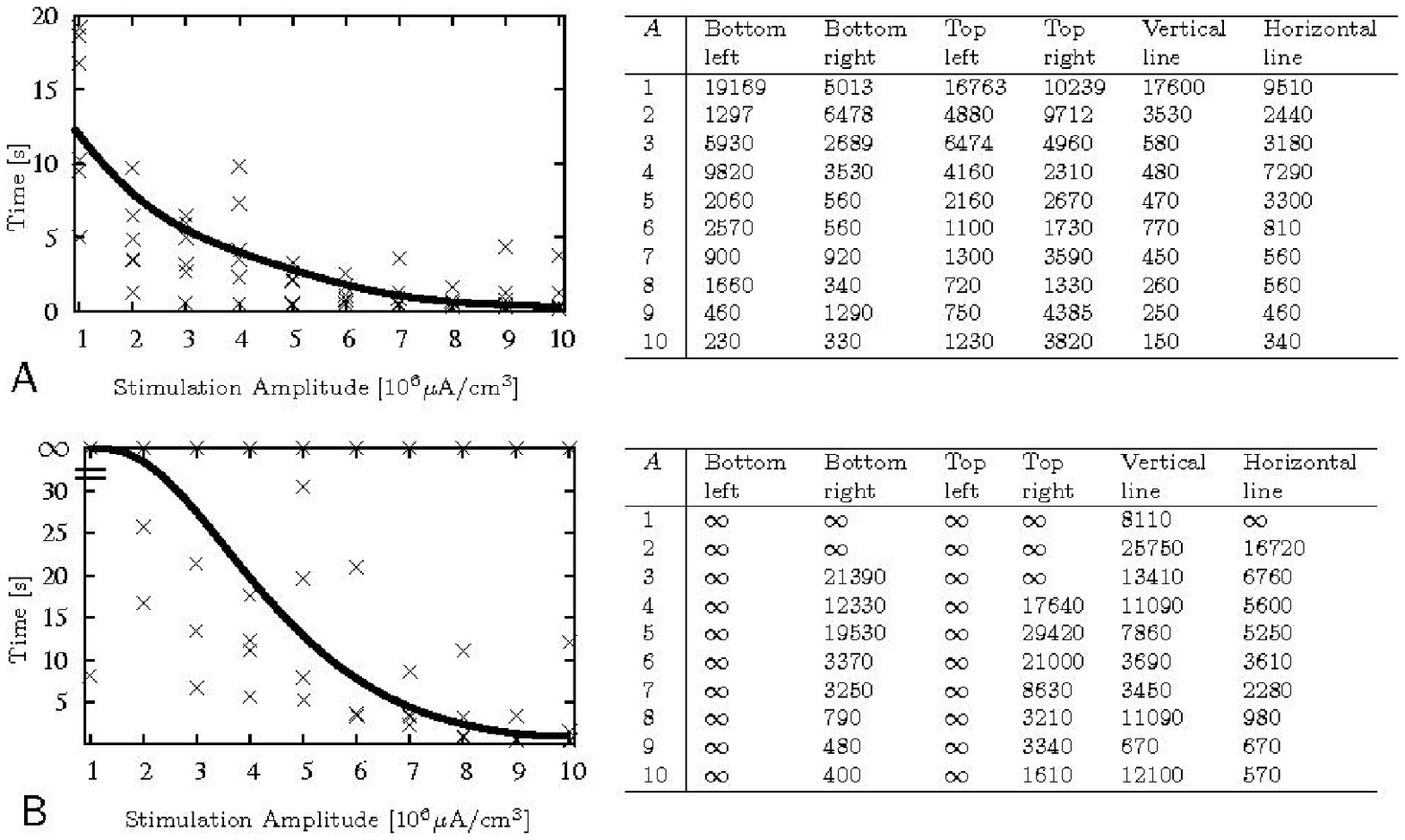}{
  Resonant drift pacing results.
  A: Meandering re-entry. 
  Left: termination time as a function of shock amplitude $A$,
  smooth curve: Bezier approximation.
  Right: the raw data, time  in $\ms$ and 
  amplitude $A$ in $10^6\,\uA/\cm^{3}$. 
  B: Same, for a stationary rotating re-entry, $\infty$: infinite loop.
}{figresults}

The time taken for termination of all re-entrant activity using
different locations of the registration electrode is shown in
\fig{figresults}A. It shows that resonant drift pacing can
successfully terminate meandering re-entrant patterns using shocks 14
times weaker than the single shock defibrillation
threshold. Termination was achieved in all simulations in this series.

Among the registration electrodes, the vertical line was the most
successful, as it produced the largest proportion of successful
terminations in the fastest time except at the smallest
amplitudes. Using a point-electrode in the top right location was the
least successful.

We conclude that the termination time depends on the mutual position
of the registration electrode, anode, cathode, initial position of the
re-entry, tissue size and fibre orientation.

For amplitude $A=1\times 10^6 \uA/\cm^{3}$, re-entry termination time
for the meandering pattern, in most cases, was longer than its
self-termination time $16000\,\ms$. Thus, we did not consider lower
values for the stimulation amplitudes.
 
\subsubsection*{Stationary rotating re-entry} \label{SR}

\Fig{figresults}B shows the times taken for annihilation of all
re-entrant waves for the different locations of the registration
electrode.  Resonant drift pacing can terminate stationary-rotating
patterns using amplitudes 18 times lower than the single shock
threshold. The combination of shock strength with the location of the
registration electrode affects the probability of success and the time
taken for termination.

\myfigure{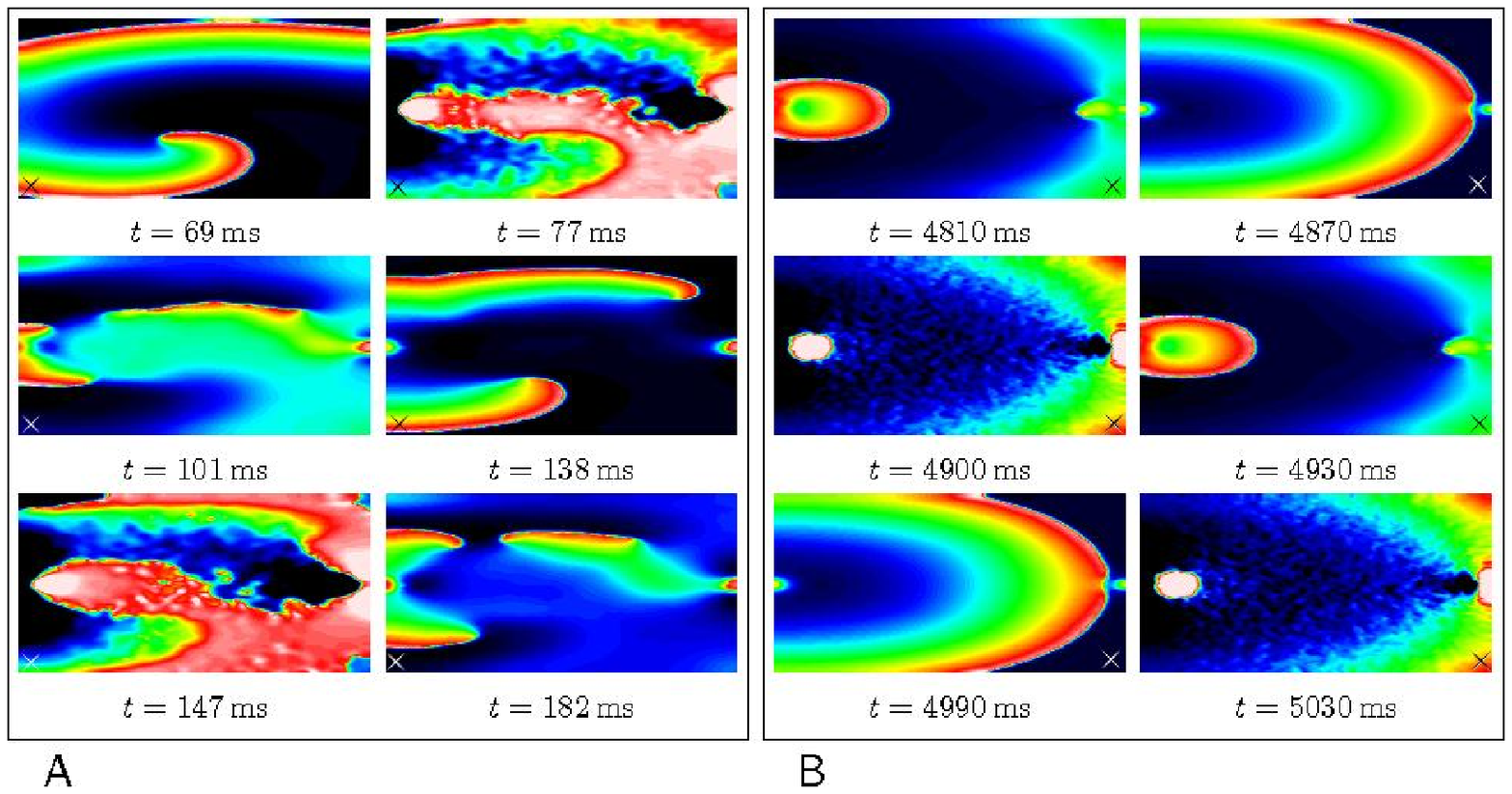}{
  Infinite loops in resonant drift pacing. 
  A: $A=10 \times 10^6\,\uA/\cm^3$. New waves with breaks are initiated
  by shocks and trigger further shocks. 
  B: $A=2 \times 10^6\,\uA/\cm^3$. New waves without breaks are initiated
  by shocks and trigger further shocks. 
}{loop}

Termination of stationary rotating re-entry was not always observed
within the $30000\,\ms$ limit.  Sometimes the algorithm gets caught in
an infinite loop (\fig{loop}), when the original re-entrant pattern
has been terminated, but shocks produce new wavefronts which trigger
further stimulation producing further wavefronts and so on.

\subsection*{Direction of drift}

The theory of the resonant drift \cite{Davydov-etal-1988} predicts
that the drift direction depends on the stimulation phase. This
dependence leads to the relationship between the drift direction and
the delay in the case of a feedback controlling forcing
\cite{Biktashev-Holden-1994,Biktashev-Holden-1995}.  A delay by a
certain fraction of a spiral's period causes a change in the drift
direction by the same fraction of $360^{\circ}$.  We have verified
that it works in \emph{this realistically anisotropic bidomain model}
as well.

\myfigure{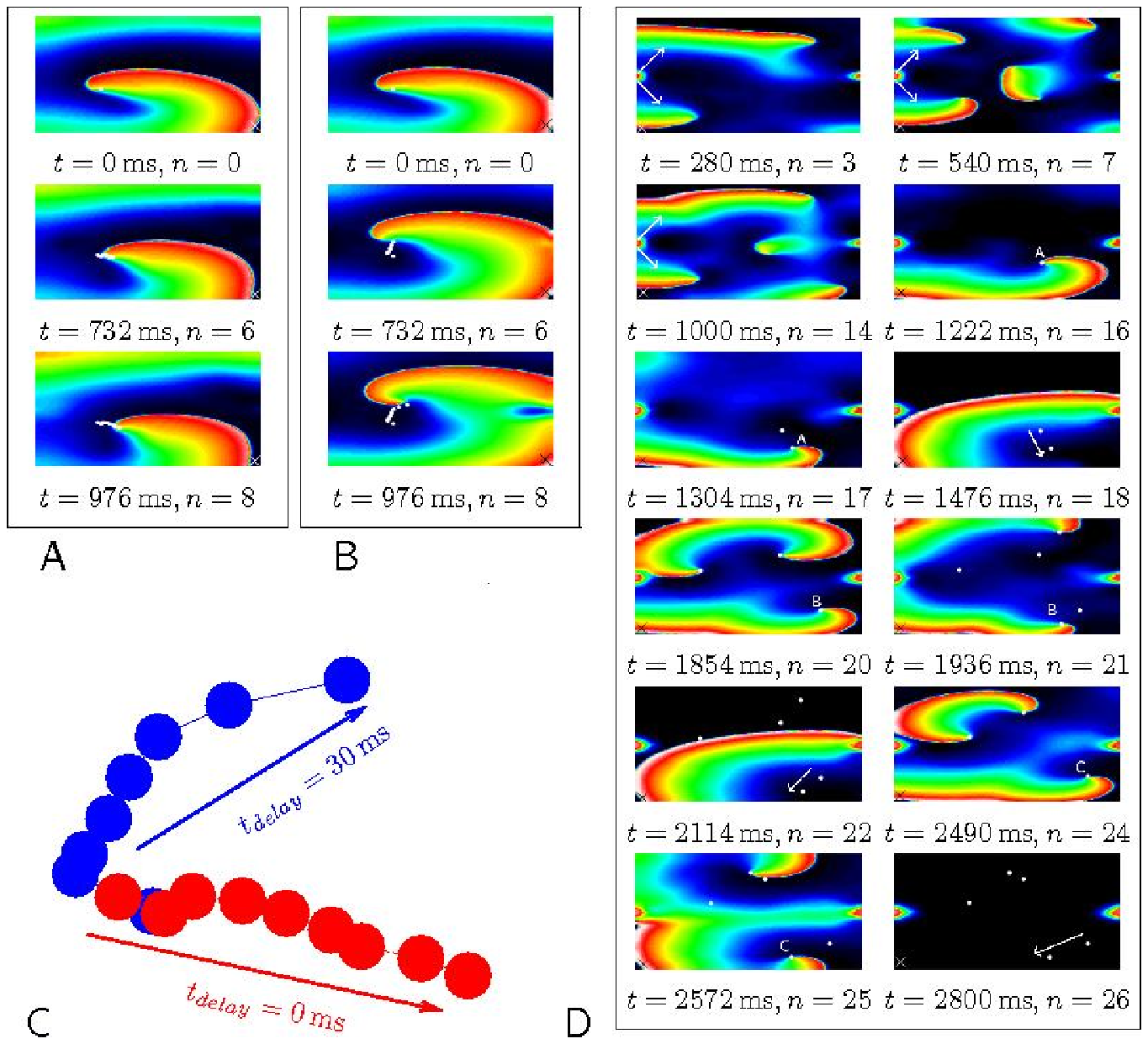}{
  Resonant drift of a meandering pattern with 
  $t_{delay} = 0\,\ms$ (A) and $t_{delay}=30\,\ms$, $A=1\times10^{6}\,\uA/\cm^3$ (B). 
  C: Tip trajectories of A, B enlarged.
  D: Effect of increasing $t_{delay}$ to $30\,\,\ms$ at
  $t=1000\,\ms$ to the infinite loop shown in 
  \fig{loop}A: the re-entry is extinguished. Arrows:
  generation of new waves and the direction of drift.
  \textit{NB}: resonant drift pacing handles multiple patterns
}{delaysuccess}

\Fig{delaysuccess}A shows fragments of two such simulations, with the
same initial conditions and different values of $t_{delay}$, $0$ and
$30\,\ms$, which is about a quarter of the re-entry period.  The
direction of the drift in these two cases differs roughly by about
$90^{\circ}$, with account of the anisotropy, in agreement with
\cite{Biktashev-Holden-1994,Biktashev-Holden-1995}.

\subsection*{Using a time delay to induce success}

We have mentioned in section \ref{SR} that resonant drift pacing may
fail through an infinite loop of wavefront eliminations and
creations. If newly created wavefronts are unbroken as in \fig{loop}B
then the re-entrant activity will vanish if the stimulation is stopped
at any time, \ie\ the success is achieved.

A genuine failure happens when newly created waves are broken and form
new re-entries, as in \fig{loop}A.  The chances of getting in such a
loop depend on the drift trajectory, which depends on the location of
the registration and stimulation electrodes and the feedback delay.
Hence such loops should be avoidable by an appropriate choice of the
electrode locations and/or of $t_{delay}$.  In a real-life situation,
such choice would require plenty of information and may be difficult
to make. Not all theoretical solutions (\eg\ optimal positions of the
electrodes) may be possible to implement in practice. Further, the
optimal parameters may vary from one arrhythmia episode to another so
difficult to predict \textit{a priori}.

Of all the mentioned parameters, $t_{delay}$ is the easiest to change,
and it may even be adjusted in real time, during pacing.  Hence we
propose that a change of the feedback delay may be used to discontinue
an infinite loop \emph{even after it has already started}.

\Fig{delaysuccess}D illustrates the feasibility of this approach.  It
is a re-run of the simulation of \fig{loop}A, with $t_{delay}$
increased to $30\,\ms$ at $t=1000\,\ms$. This broke the loop and
terminated the re-entry.


\section*{Discussion}

\subsection*{Summary of results: 
Resonant drift pacing works in a realistic model of electric field action on cardiac tissue}

This simulation study presents results which can be used as
experimentally testable hypotheses. It has been based on the
understanding of the mechanism by which external electric current
affects excitation and propagation via non-uniformity of the electric
field \emph{and} the heterogeneity of the tissue conductivities.  This
mechanism implies both a strong action near electrodes due to the
electric field inhomogeneity, and a weaker, but far reaching action in
the bulk of the tissue due to omnipresent tissue heterogeneities. This
model has been used before to simulate single-shock defibrillation
\cite{Plank-etal-2005}. Here we have demonstrated that the far-field
action of the electric current can cause, under repetitive
stimulation, a drift and elimination of the re-entrant sources. In our
simulations, resonant drift pacing can eliminate re-entry, including
multiple re-entries, with high probability and within acceptable time,
at amplitudes much lower than single shock defibrillation.  If we
allow 10\,s for low-voltage re-entry termination, as is the case for
antitachycardia pacing, the required shock strength with resonant
drift pacing is 12 to 15 times smaller than with a single shock.

Our simulations have shown that a major possible obstacle to
elimination of re-entrant sources is the possibility that the electric
shocks create new re-entrant waves while eliminating existing ones,
which may lead to infinite loops of annihilation and creation. The
creation of the new sources occurs near the shock electrodes where the
electric field is highly inhomogeneous, and the 12 to 15-fold decrease
in defibrillation threshold is observed \emph{notwithstanding} this
effect. The highly inhomogeneous electric field is created by point
electrodes, so bigger electrodes, which create more homogeneous
fields, should perform even better.

Our simulations have also demonstrated yet another way to overcome
infinite annihilation-creation loops. This is due to the dependence of
the direction of resonant drift on the time delay in the feedback
loop.  This dependence has been predicted and observed in isotropic
monodomain models with simplified description of the electric field
action \cite{%
  Davydov-etal-1988,%
  Biktashev-Holden-1994,%
  Biktashev-Holden-1995,%
  Biktashev-Holden-1995a,%
  Biktashev-Holden-1996%
}.  Here we demonstrated that this dependence is still observed in the
present more realistic model and, moreover, it can be used to
discontinue an infinite annihilation-creation loop, via a change in
the feedback delay once such a loop has been detected.
  
Although we used atrial tissue kinetics, there are indications that
the exact sort of excitable kinetics is not too important for the
properties of resonant drift
\cite{Biktashev-Holden-1995a,Biktashev-Holden-1996}.  So we expect
that our present results may be interesting for ventricular
fibrillation, too.

\subsection*{Comparison with previous research}

\paragraph{Anti-tachycardia pacing}
In clinical practice, antitachycardia pacing is as efficient and as
safe as single shock defibrillation, even for fast ventricular
tachycardias with up to 250~beats/min~\cite{ATP2}.  The exact
mechanisms underlying success or failure of antitachycardia pacing are
widely unexplored.  The traditional understanding implies stimulation
which is faster than the anatomical re-entry and which engages all the
larger area of tissue until it reaches an ``isthmus'' of the re-entry
and blocks it. It is conceivable, however, that in some cases
antitachycardia pacing may be unwittingly applied to a functional
rather than anatomical re-entry and then its mechanism could be the
resonant drift. In such cases, feedback-controlled resonant drift
pacing will do a better job than pacing with prescribed frequency used
in antitachycardia pacing.  Although the probability of
antitachycardia pacing success is high, when it fails a single strong
shock has to be applied.

\paragraph{Unpinning}
It has been speculated that anatomically or functionally anchored
ventricular tachycardias are less likely to be terminated by
antitachycardia pacing.  For these cases, it has been suggested that a
weak shock, with an amplitude in the range used in this study, can be
applied to \emph{unpin} the re-entry from the obstacle by a virtual
electrode polarization mechanism
\cite{ripplinger06:_unpinning,tung06:_perturb}.  Such a stimulus must
be correctly timed, which can be achieved by synchronizing it to the
signal from a registration electrode, with a correctly chosen delay
with respect to that signal, which in practice may require scanning
through possible delays.  The resonant drift pacing also requires
synchronization with a registered signal, and we have seen that change
in the delay during the course of pacing may also be beneficial.  This
makes the unpinning protocol operationally very similar to our
resonant drift pacing protocol, and in any particular case it may not
be possible to say with certainty which mechanism has worked.

\paragraph{Experiments with feedback-controlled pacing}

Pak \etal\ \cite{Pak} eliminated ventricular fibrillation in rabbit by
multisite pacing synchronized with optical signal from a fixed
reference site.  Their stimulation protocol was similar to resonant
drift pacing considered here, except they used more pacing electrodes
and aimed to deliver shocks only when pacing site was in an excitable
gap, under the implicit assumption that shock delivered to a site in
an absolute refractory state could not possibly affect the re-entry,
i.e. \textit{a priori} ignoring far-field effects.  Despite this
artificial self-limitation, they were successful at defibrillating
with shock strengths an order of magnitude lower than single shock
defibrillation and with better success rate than overdrive and
high-frequency pacing. It was argued in \cite{Pak} that earlier
mathematical modeling of resonant drift only demonstrated termination
of a single re-entrant wave (see, however,
\cite{Biktashev-Holden-1995}) and their experiments did not show
evidence of extinction of re-entries on ventricular borders.  Our
present simulations indicate that
\begin{itemize}
\item far-field effects can make it worthwhile issuing shocks
  regardless of the tissue state at the pacing site,
\item the resonant drift pacing can work for multiple re-entrant waves
  (providing that the difficulties of new wavefronts being initiated
  by the shocks are overcome), and
\item annihilation at the boundaries is not a necessary feature of
  resonant drift elimination, as re-entry sources can annihilate with
  each other.
\end{itemize}
Thus, we find the results by Pak \etal, contrary to their own
interpretation, as strong experimental evidence of the resonant drift
pacing being an efficient method of low voltage defibrillation.

\subsection*{Clinical implications}

Resonant drift pacing may present an alternative or a supplement for
existing (single shock defibrillation, synchronized cardioversion,
antitachycardia pacing, overdrive pacing, and high-frequency pacing)
and proposed (unpinning) therapies. The theoretical mechanism
underlying this protocol is for functionally determined re-entries,
\ie\ probably for higher frequency tachycardias and fibrillations. Due
to similarity between resonant drift and unpinning pacing protocols,
the two protocols may be possible to combine into one, which would
work via unpinning for pinned (monomorphic tachycardias), and via
resonant drift for unpinned re-entrant waves (polymorphic tachycardias
and fibrillations).

There is evidence that some atrial arrhythmias are not re-entrant, but
are due to rapid ectopic focal activity \cite{Nattel-2002}. The
proposed method might work on such arrhythmias via a completely
different mechanism, say overdrive suppression, but this can only be
clear after further studies.

As the study presented here is purely theoretical, the key question is
whether the low-energy approach to defibrillation considered here
could work in real heart. Clearly this question can only be answered
by experimental studies, and there is clear scope to advance the field
by interaction between modeling studies such as this one and
experimental studies.

\subsection*{Study limitations}

Apart from inevitable limitations inherent in mathematical modeling
compared to experimental study, which are due to limited current
knowledge, we have made a number of simplifications.  Our model tissue
was essentially two- rather than three-dimensional.  Its geometry was
not realistic (rectangular shape).  The model tissue lacked any
macroscopic inhomogeneities, such as trasmural, center-periphery or
base-apex gradient of excitability properties, or variations of the
conductivity tensor.  The position and geometry of the electrodes was
arbitrary.  Values of some of the parameters lack reliable
experimental foundation, including some of the most important ones,
such as $F$, the intracellular conductivity fluctuations amplitude.
The spatially uncorrelated structure of conductivity fluctuations is
an idealization.  These limitations can be lifted as more experimental
data become available and via further, more detailed, simulation
studies.

Resonant drift pacing may potentially work for ventricular tachycardia
and fibrillation too.  However, the ventricles have thick walls and
overall more complicated anatomy, which may pose extra difficulties
for defibrillation. So any extrapolation of our present results to
ventricles should be thought of with caution, bearing in mind possible
complications. There are modeling studies suggesting that at least
some of the new difficulties arising in resonant pacing of
three-dimensional re-entries can be successfully overcome (see
\cite{Morgan-etal-2008} and references therein), however those were in
overly simplified models of exciable media, and more realistic
computational studies are still desirable.

\subsection*{Conclusion}

Our results show that in a bidomain model of cardiac tissue, with
microscopic conductivity fluctuations, re-entrant waves can be
annihilated using feedback-controlled repetitive stimulations by
inducing resonant drift and directing the re-entry towards an
inexcitable boundary. If associated difficulties are overcome,
termination using this approach is achieved, with high probability and
within acceptable time, at a fraction of the conventional shock
strength. The direction of the drift can be managed by choosing the
location of the electrodes and the time delay of the shock
application.

Difficulties occur due to new wavefronts being initiated from the
shock electrodes. However, numerical simulations allow a detailed
insight into this problem and can be used as a tool to suggest
solutions.

Our findings present experimentally testable hypotheses for what we
expect to observe in real cardiac tissue. There is scope to advance
the field of low-energy defibrillation by interaction between modeling
studies such as this and experimental studies.

\section*{Acknowledgements}
This study was supported in part by
EPSRC grant EP/D500338/1,
Royal Society grant 2005/R4,
Liverpool University Research Development Fund grant 4431 
and
Austrian Science Fund FWF grant F3210-N18. 


\bibliography{rd2b}

\end{document}